\documentclass[preprint,showpacs]{revtex4} 

\usepackage{amssymb}
\usepackage[dvips]{epsfig}
\usepackage{subfigure}
\usepackage{amsmath}

\newcommand{\scl}{0.2} 

\begin{document}

\title{Dynamic power balance for nonlinear waves \\ in unbalanced gain and loss landscapes}

\author{Yannis Kominis}
\affiliation{School of Applied Mathematical and Physical Science, National Technical University of Athens, Zographou GR-15773, Greece}
\begin{abstract}
The presence of losses in nonlinear photonic structures is a crucial issue for modern applications. Active parts are introduced for wave power compensation resulting in unbalanced gain and loss landscapes where localized beam propagation is, in general, dynamically unstable.  Here we provide generic sufficient conditions for the relation between the gain-loss and the refractive index profiles in order to ensure efficient wave trapping and stable propagation for a wide range of beam launching conditions such as initial power, angle of incidence and position. The stability is a consequence of an underlying dynamic power balance mechanism related to a conserved quantity of wave dynamics.
\end{abstract}
\maketitle

Modern photonic applications utilize the combination of nonlinearity and inhomogeneity in order to provide advanced functionality in properly engineered metamaterials and metadevices. \cite{ZhKi_12, MiMi_15} These structures have the form of multilayered media consisted of materials such as ordinary dielectrics as well as metals and graphene. The presence of metals results in plasmonic excitations that can boost nonlinear effects due to high field values and small mode volume \cite{KaZa_12} whereas the presence of graphene layers is accompanied with very strong Kerr nonlinearities, \cite{ZhVi_12} both resulting in the formation of self-localized modes. \cite{BoCo_86, FeOr_07, AuAn_13, BlSm_15} The nonlinearity plays a crucial role in functionality related to dynamic and all optical light control through wave-material and wave-wave interactions. However, both ordinary dielectrics and metal or graphene layers introduce significant losses that can hamper the nonlinear functionality of these structures by restricting the wave propagation to small distances. \cite{TaKo_12} This crucial drawback of the respective photonic structures necessitates the utilization of active parts (hot-spots) in the form of doped and pumped dielectrics in order to provide the necessary gain for loss compensation, \cite{NoZh_08, SaSh_13, MaRo_15} introducing an inhomogeneous gain-loss landscape. A similar type of nonconservative inhomogeneity also appears in applications related to soliton-forming laser cavities. \cite{GrAk_12} In all these cases the formation and robust propagation of a self-localized mode is determined by both the diffraction-nonlinearity and the loss-gain balance, which cannot be considered separately.

From an engineering aspect, even the presence of spatially homogeneous gain and loss in an optical lattice can significantly enrich soliton dynamics providing soliton routing and acceleration functionalities, \cite{KoDr_12, KoPa_12} in addition to the numerous applications related to conservative lattices that include the formation of solitons, surface waves and defect modes. \cite{TrTo_01, KiAg_03} Moreover, the appropriate design of gain and loss inhomogeneity provides another degree of freedom for wave manipulation. \cite{HeMi_13, Ko_15} The symmetry properties of the inhomogeneity profiles have been shown to play a crucial role on the system features. It has been shown that, for the case of $\mathcal{PT}$ symmetry, the system has a real spectrum and supports a continuous family of solitons. \cite{MuMa_08, RuMa_10} Other types of symmetries that restrict, not the profiles of the refractive index and the gain-loss inhomogeneity as in the $\mathcal{PT}$ case, but their mutual relation, have been also shown to support such continuous soliton families, \cite{Ko_15, TsAl_14, KoZe_14} in contrast to the common case of dissipative solitons where, in general, only isolated solitons exist.

A solitary wave can propagate at a fixed transverse position of a planar structure near the interface between a gain- and a loss-region, where a static power balance condition is satisfied. However, any deviation from this specific position or from a zero angle of incidence can lead to continuous power increasing or decreasing, resulting in an unstable behavior, as in the case of stationary solitons pinned to hot-spots. \cite{LaMa_09, TsMa_10, TsMa_11, ZeKa_11, MaDi_12} The utilization of spatial modulations of the linear or the nonlinear refractive index has been proposed \cite{TsTa_12, MaDr_13} for introducing effective potential wells resulting to wave trapping in the specific position and preventing large excursions within the two regions around the fixed position. Even in such cases, wave oscillations around the balance position, can be unstable when the gain and loss of the interfaced parts are unbalanced, as in the most typical case where narrow hot-spots with high gain are utilized in order to compensate for more extended parts with relatively small losses. The instability arises from the fact that the dynamic power balance of the wave depends on the extent of the oscillations in the two parts, since the wave amplification and attenuation in the two phases of the oscillation are not equal in general. Therefore, an appropriate refractive index modulation has to take into account the gain and loss profile, in order to ensure a dynamic balance of gain and loss and a stable wave propagation. 

In the following, we present generic efficient conditions for the relation between the gain-loss and the refractive index profiles allowing, not only for stable stationary propagation at a specific point with a zero angle of incidence, but also for dynamic power balance for localized waves with a wide range of positions and angles of incidence, that are applicable to any type of planar photonic structure that may have unbalanced gain and loss properties.

\section*{Model and Method}
Nonlinear wave propagation in a transversely inhomogeneous planar photonic structure is described by the NonLinear Schr\"odinger Equation (NLSE)
\begin{equation}
iu_z+u_{xx}+\left[V(x)-iW(x)\right]u+2|u|^2u=0 \label{NLSE}
\end{equation}
where $u$ is the normalized electric field envelope, $z$ and $x$ are the normalized longitudinal and transverse dimensions, and $V(x)$, $W(x)$ are the transverse refractive index and gain-loss profiles, respectively. For spatially localized (solitary) waves, we can define the useful quantities corresponding to the wave mass $m=\int|u|^2dx$ and momentum $p=i\int(uu_x^*-u_xu^*)dx$. In the absence of inhomogeneity ($V=W=0$), $m$ and $p$ are conserved, whereas under the presence of inhomogeneity they vary as
\begin{eqnarray} 
\frac{dm}{dz}&=&\Gamma(x_0) \label{dmdz} \\
m\frac{dv}{dz}&=&-\frac{\partial U_{eff}(x_0)}{\partial x_0} \label{dpdz}
\end{eqnarray}
where
\begin{eqnarray}
\Gamma(x_0) &=& 2\int_{-\infty}^{+\infty}|u(x-x_0)|^2W(x)dx \label{Gamma} \\
U_{eff}(x_0)&=&-2\int_{-\infty}^{+\infty}|u(x-x_0)|^2V(x)dx \label{U_eff}
\end{eqnarray}
are the mass variation rate and the effective potential, respectively, and $x_0$ is the wave center varying as $dx_0/dz=p/m \equiv v$ with the velocity $v$ corresponding to the propagation angle. Therefore, the solitary wave propagates as an effective particle with mass and momentum variations depending on the nonconservative [$W(x)$] and the conservative [$V(x)$] part of the inhomogeneity, respectively. Wave propagation dynamics are described in the three-dimensional space $(x_0,v,m)$. It can be readily shown \cite{Ko_15} that, under the condition
\begin{equation}
\frac{\partial V(x)}{\partial x}=C W(x), \label{condition}
\end{equation}  
the existence of an exact invariant of the motion, given by $K=C\ln m+v$, restricts the wave dynamics in a two-dimensional surface. This is a general property of any type of solitary wave in the presence of inhomogeneities of arbitrary profile and magnitude. The condition (\ref{condition}) ensures the static power balance for a stationary solitary wave located at a fixed point $\Gamma(x_0)=0$ at the vicinity of the interface between a lossy and an amplifying part. Moreover, it is a stronger condition, sufficient for the dynamic power balance of solitary waves with nonzero angles of incidence and positions deviating from the fixed point that undergo stable oscillations, as we show in the following. Notice that the condition (\ref{condition}) is qualitatively different from the $\mathcal{PT}$ symmetry condition, since it does not imply any restriction on the symmetry properties of the nonconservative [$W(x)$] and the conservative [$V(x)$] part of the inhomogeneity, but only a mutual relation of their profiles; therefore, it is also applied in non-symmetric profiles.  Under condition (\ref{condition}) when $V(x)$ is even, $W(x)$ is odd (and vice versa) as in $\mathcal{PT}$ symmetric configurations. However, the $\mathcal{PT}$ symmetry condition suggests only the existence of a fixed point and does not ensure its stability.

In the following, we exploit the consequences of this condition with respect to the dynamic power balance for solitary waves in a wide variety of planar structures, and prove that Eq. (\ref{condition}) serves as a generic sufficient condition for stable wave propagation in accordingly designed photonic structures.

\section*{Results and Discussion}
We focus on multilayer photonic structures with piecewise constant gain and loss profiles. According to Eq.(\ref{condition}), the linear refractive index profile is a piecewise linear function, so that  
\begin{equation}
W(x)=\sum_i \Pi_i(x),  \hspace{2em} V(x)=\sum_i \Lambda_i(x) \label{structure}
\end{equation}
with 
\begin{eqnarray}
\Pi_i(x)&=& 
\left\{ \begin{array}{ll}
		a_i, & x_{i,1}<x<x_{i,2} \\
		0, & \mbox{elsewhere}
		\end{array}
\right. \\
\Lambda_i(x)&=&
\left\{ \begin{array}{ll}
		c_ix+d_i, & x_{i,1}<x<x_{i,2} \\
		0, & \mbox{elsewhere}
		\end{array}
\right.
\end{eqnarray}

The dynamics of solitary wave propagation in such structures is determined by Eqs. (\ref{dmdz}), (\ref{dpdz}). Without loss of generality, in order to simplify our analysis and provide intuitive understanding, we consider inhomogeneities of relatively small amplitude so that we can obtain closed form equations for the Eqs. (\ref{Gamma})-(\ref{U_eff}) by utilizing the soliton solution of the homogeneous NLSE $u=\eta \mbox{sech} [\eta (x-x_0)] \exp [i(v/2)x+i(\eta^2-v^2/4)z]$ in the calculation of the respective integrals, resulting in
\begin{eqnarray}
\Gamma(x_0) &=& m\sum_i \left[ \pi_i\left(x_{i,2}\right) - \pi_i\left(x_{i,1}\right) \right] \\
U_{eff}(x_0)&=&-2m\sum_i \left[ \lambda_i\left(x_{i,2}\right) - \lambda_i\left(x_{i,1}\right) \right]
\end{eqnarray}

\begin{eqnarray}
\pi_i(x)&=& a_i \tanh\frac{m}{2}(x-x_0) \\
\lambda_i(x)&=& c_i(x-x_0)-\frac{c_i x +d_i}{e^{m(x-x_0)}+1}-\frac{c_i}{m}\ln\left(e^{m(x-x_0)}+1\right)
\end{eqnarray}
with $m=2\eta$.  

\begin{figure}[h]
   \begin{center}
  	   \subfigure[]{\scalebox{\scl}{\includegraphics{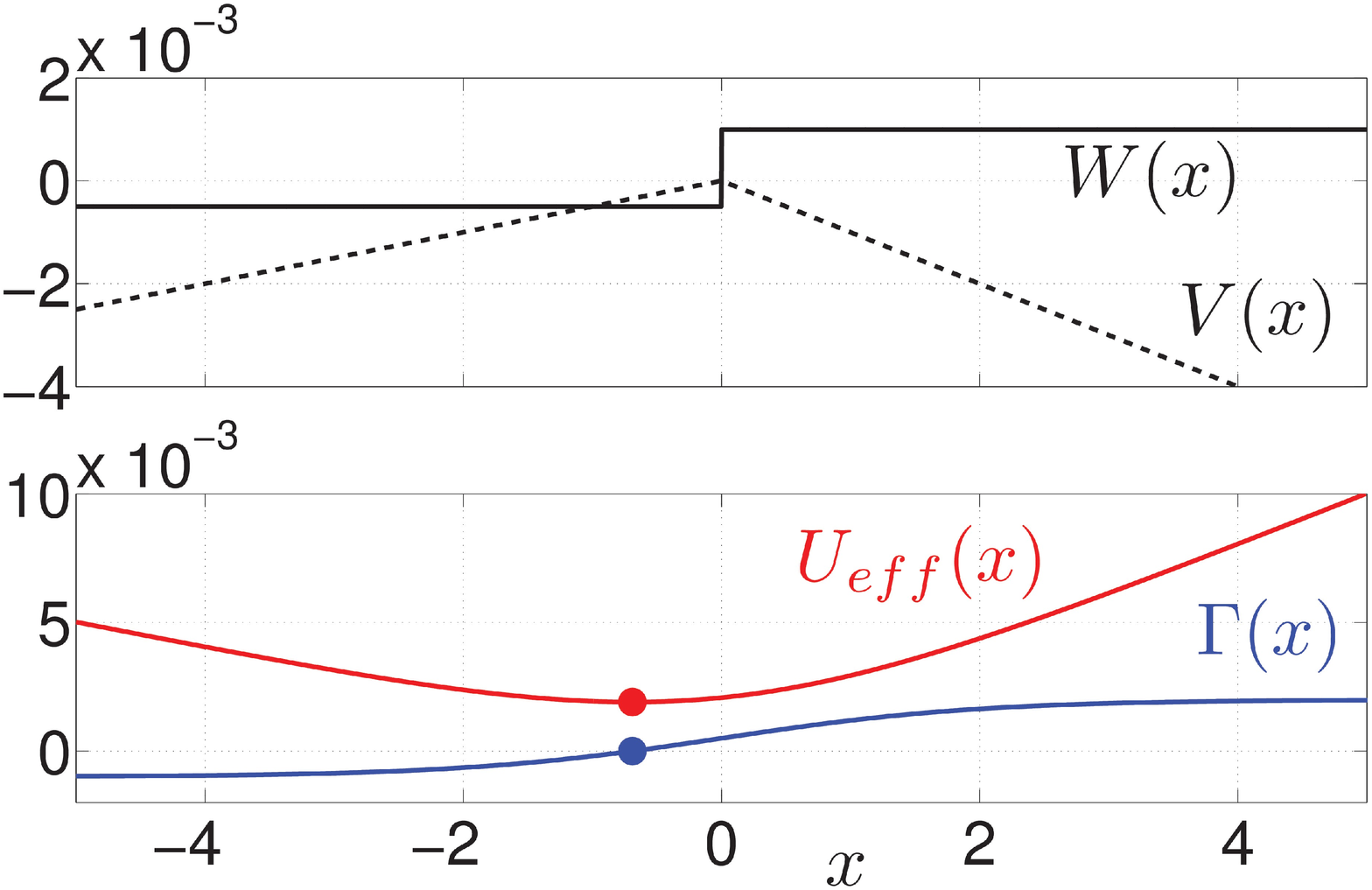}}}
	   \subfigure[]{\scalebox{\scl}{\includegraphics{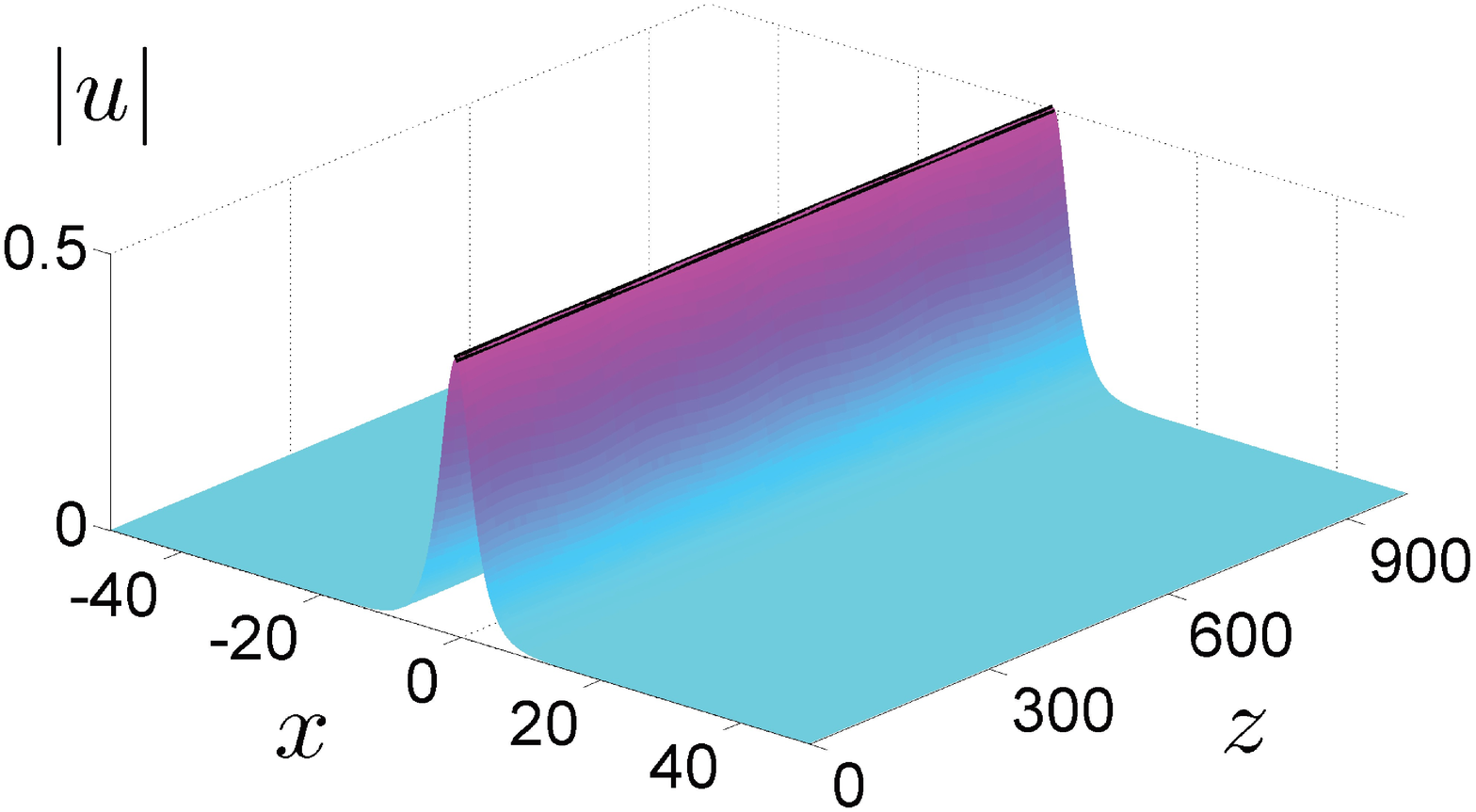}}}\\
	   \subfigure[]{\scalebox{\scl}{\includegraphics{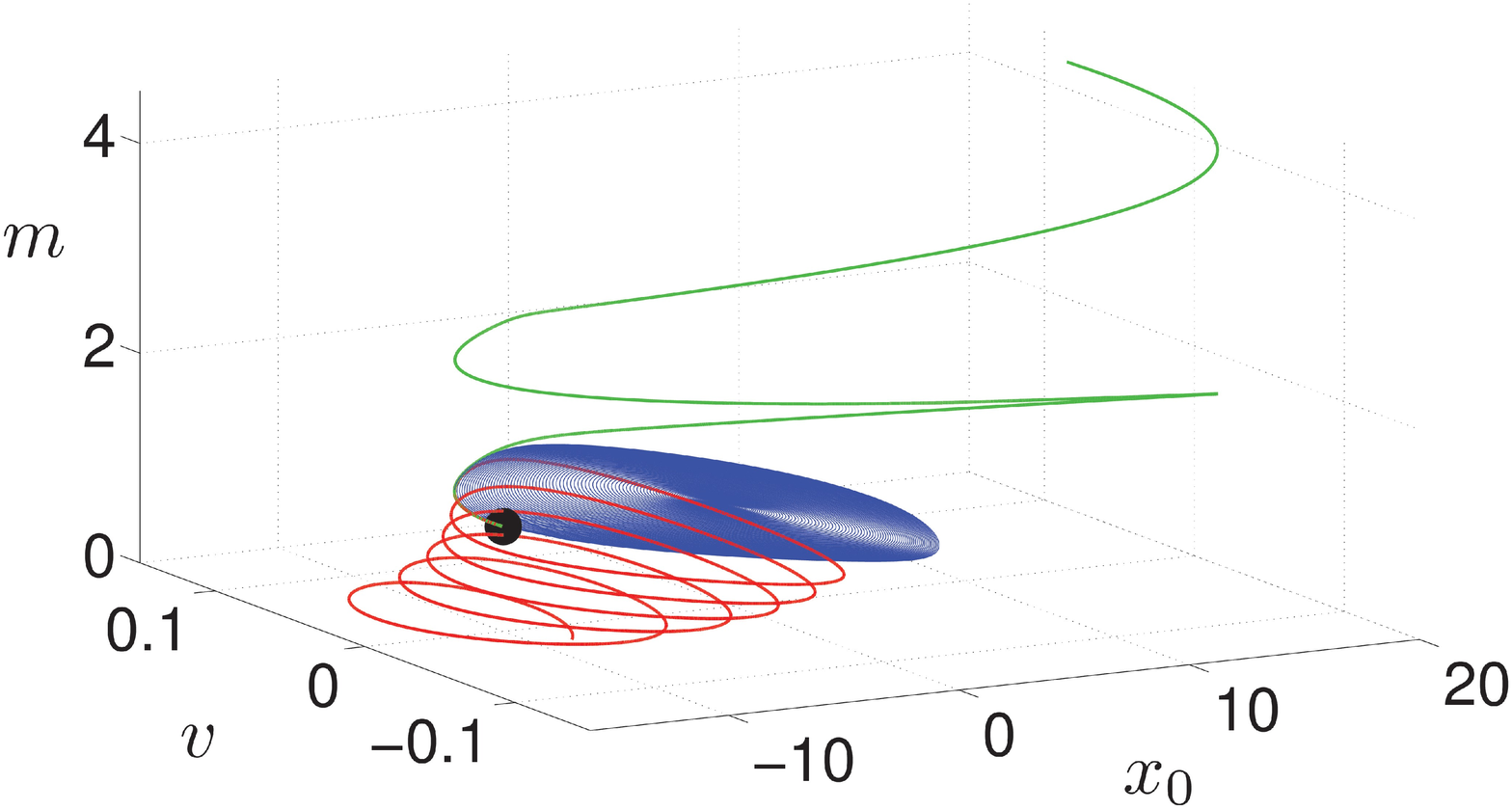}}}
	   \subfigure[]{\scalebox{\scl}{\includegraphics{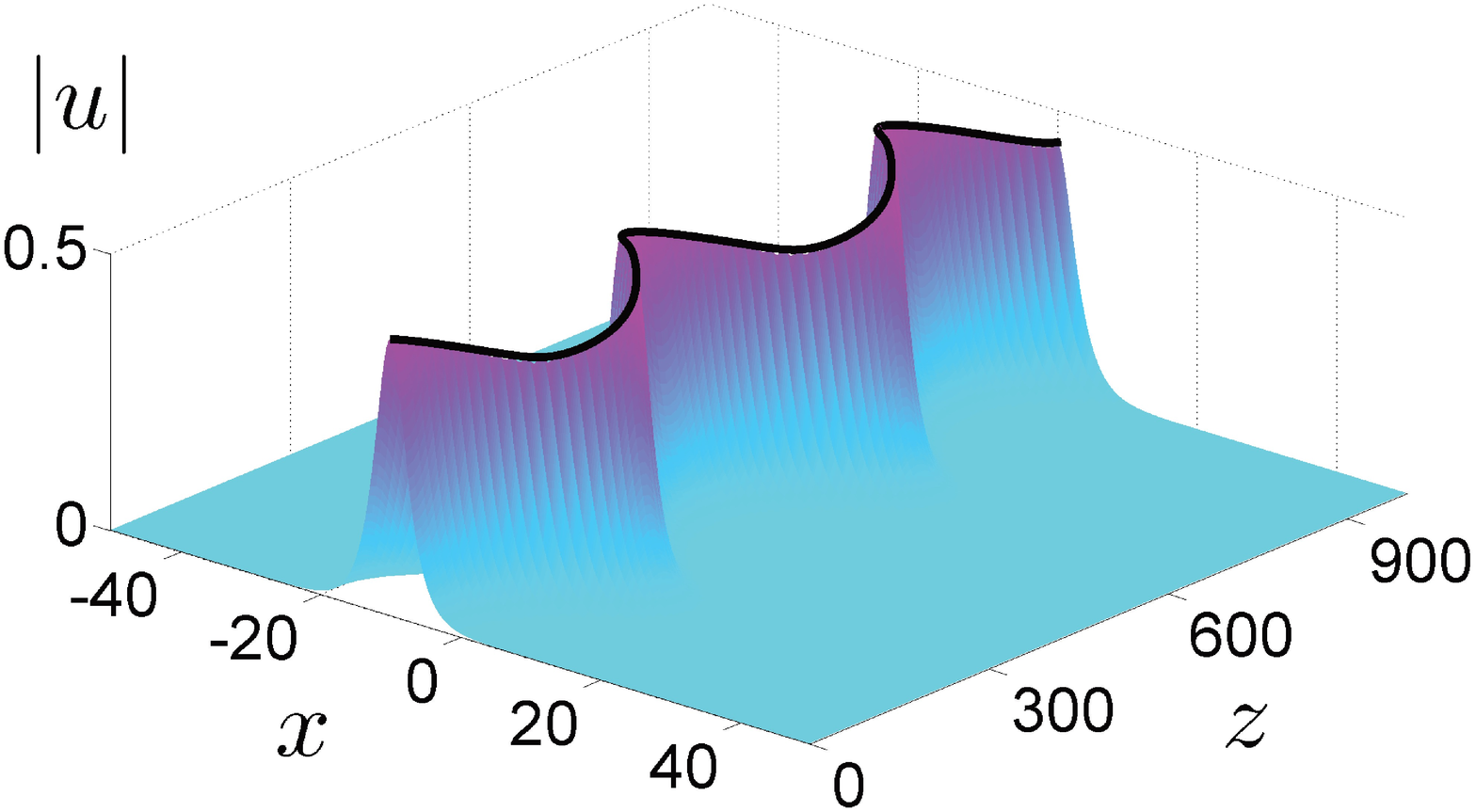}}}\\
	   \subfigure[]{\scalebox{\scl}{\includegraphics{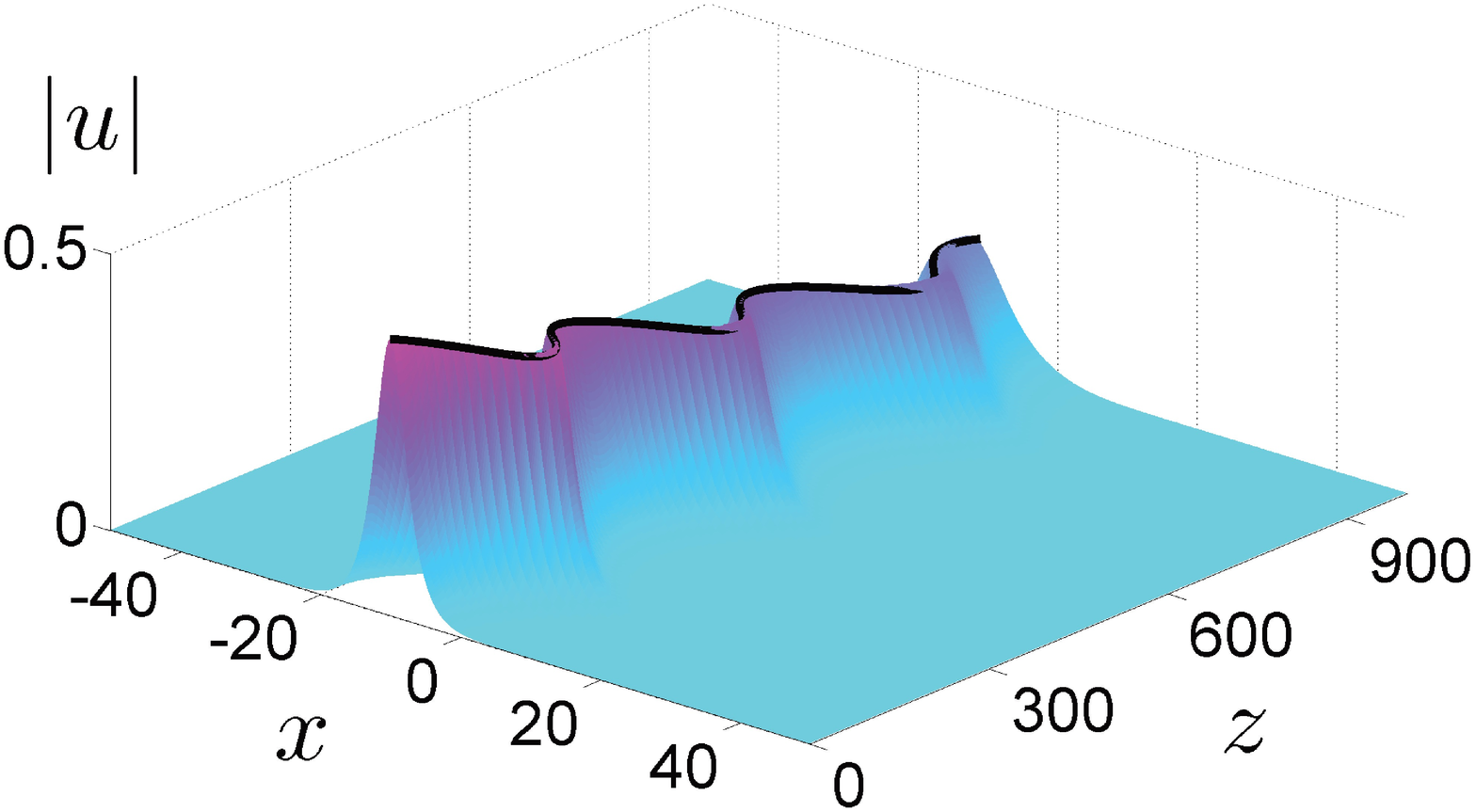}}}
	   \subfigure[]{\scalebox{\scl}{\includegraphics{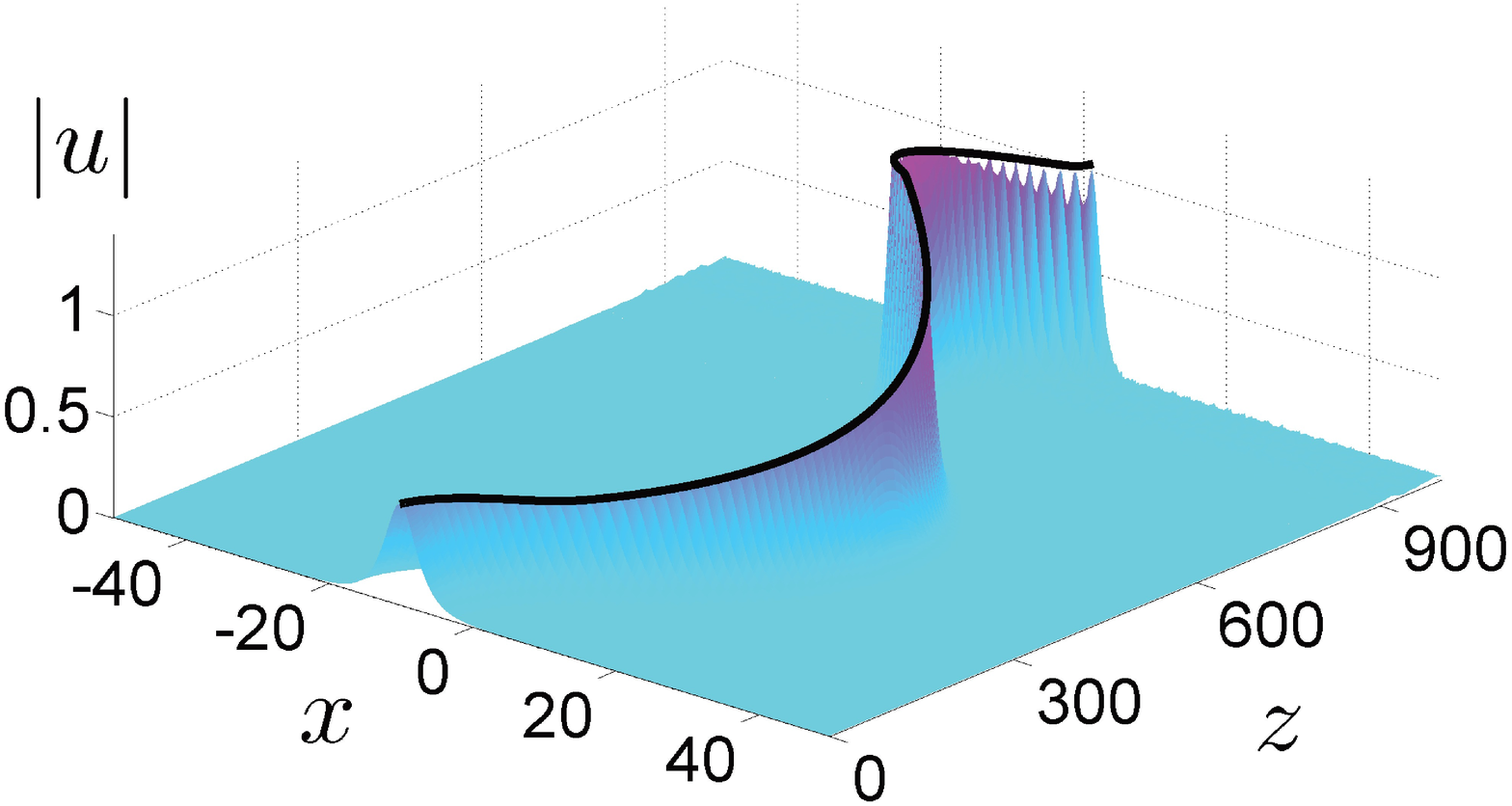}}}
\caption{Single interface between a lossy ($a_1=-0.0005$) and an amplifying ($a_2=0.001$) region. (a) Gain-loss $W(x)$ and refractive index $V(x)$ profiles [top]; Effective potential $U_{eff}(x_0)$ and mass variation rate $\Gamma(x_0)$ for a soliton of mass $m=1$ for a refractive index profile fulfilling the condition (\ref{condition}) with $C=-1$ ($c_i=-a_i, i=1,2$) [bottom]. (b) Stationary propagation for initial soliton position $x_0=-0.69$ corresponding to the fixed point depicted by a thick dot in (a). (c) Phase space orbits of a soliton with initial position $x_0=-10$  under dynamic balance conditions $c_i=-a_i, i=1,2$ (blue), and for unbalanced cases with $c_1=-a_1, c_2=-7a_2$ (red), $c_1=-a_1, c_2=-0.25a_2$ (green). (d),(e),(f) Soliton propagation for conditions corresponding to the three orbits shown in (c). The thick black lines depict results from the effective particle model.}

    \end{center}
\end{figure}

First, we consider a structure consisting of two interfaced semi-infinite parts with unequal gain and loss coeffecients and linear refractive indices profiles fulfilling the condition (\ref{condition}) with $C=-1$ as shown in Fig. 1(a). The effective potential has a local minimum, corresponding to a fixed point, in the vicinity of the interface with its exact poisition depending on the soliton mass. For a soliton of mass $m=1$ the fixed point is located at $x_0=-0.69$. Stable propagation of a stationary soliton with initial position at the fixed point is shown in Fig. 1(b). The fulfillment of the condition (\ref{condition}) results in refractive index slopes appropriately defined in terms of the gain and loss coefficients in each part. In terms of soliton dynamics, the direct consequence of the condition is that the trapping potential is such that no continuous mass increase or decrease takes place as the travelling distance in each region is such that the soliton spends less time in the high gain region than in the low loss region. In fact, this dynamic power balance mechanism results in asymptotic evolution to the stable fixed point (attractor). As shown in Fig. 1(c), the effective particle orbit for a soliton initially located at $x_0=-10$ evolves to the fixed point, while remaining in the aforementioned two-dimensional surface of the phase space. The rate of convergence to the fixed point orbit increases with the magnitude difference between the gain and loss coefficients and for the specific case is quite small as shown in Fig. 1(d). Notice that the period of oscillations scales with $|C|^{-1/2}$. The importance of the condition (\ref{condition}) can be shown in comparison to the cases where it is not fulfilled, resulting in either continuous mass decreasing or increasing as shown in Fig. 1(e) and (f), respectively, and unbounded phase space orbits [Fig. 1(c)].

\begin{figure}[h]
   \begin{center}
  	   \subfigure[]{\scalebox{\scl}{\includegraphics{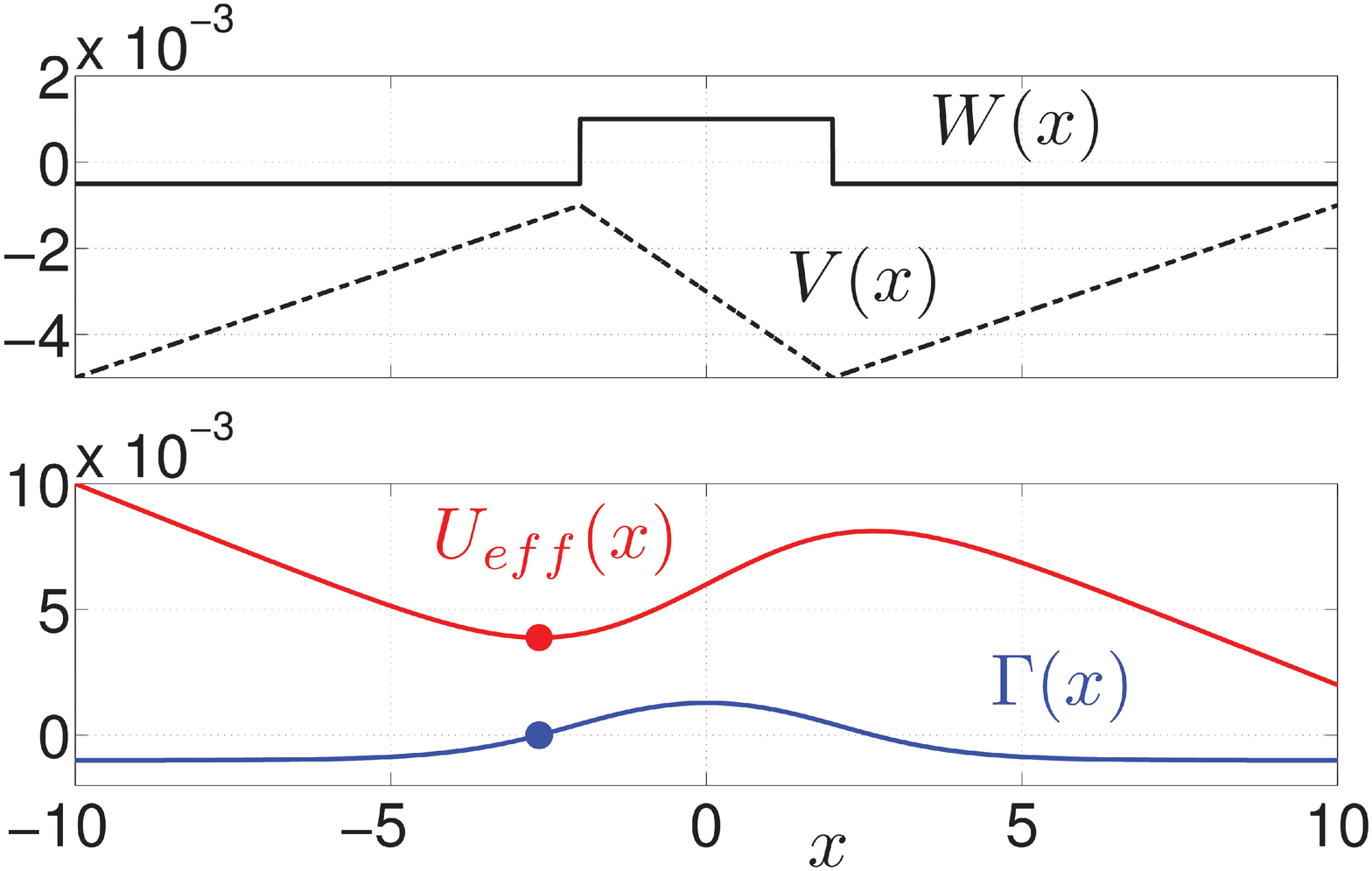}}}
	   \subfigure[]{\scalebox{\scl}{\includegraphics{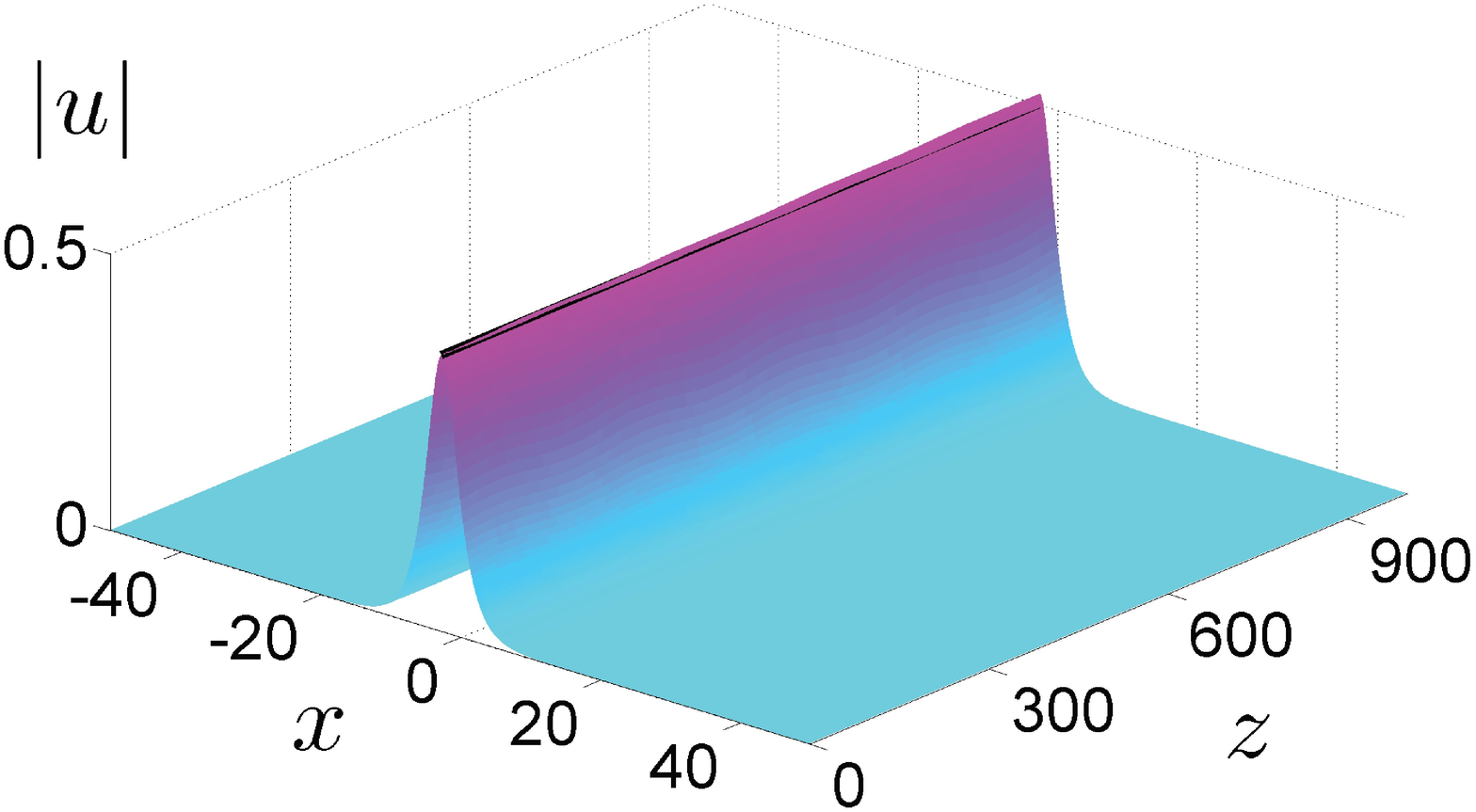}}}\\
	   \subfigure[]{\scalebox{\scl}{\includegraphics{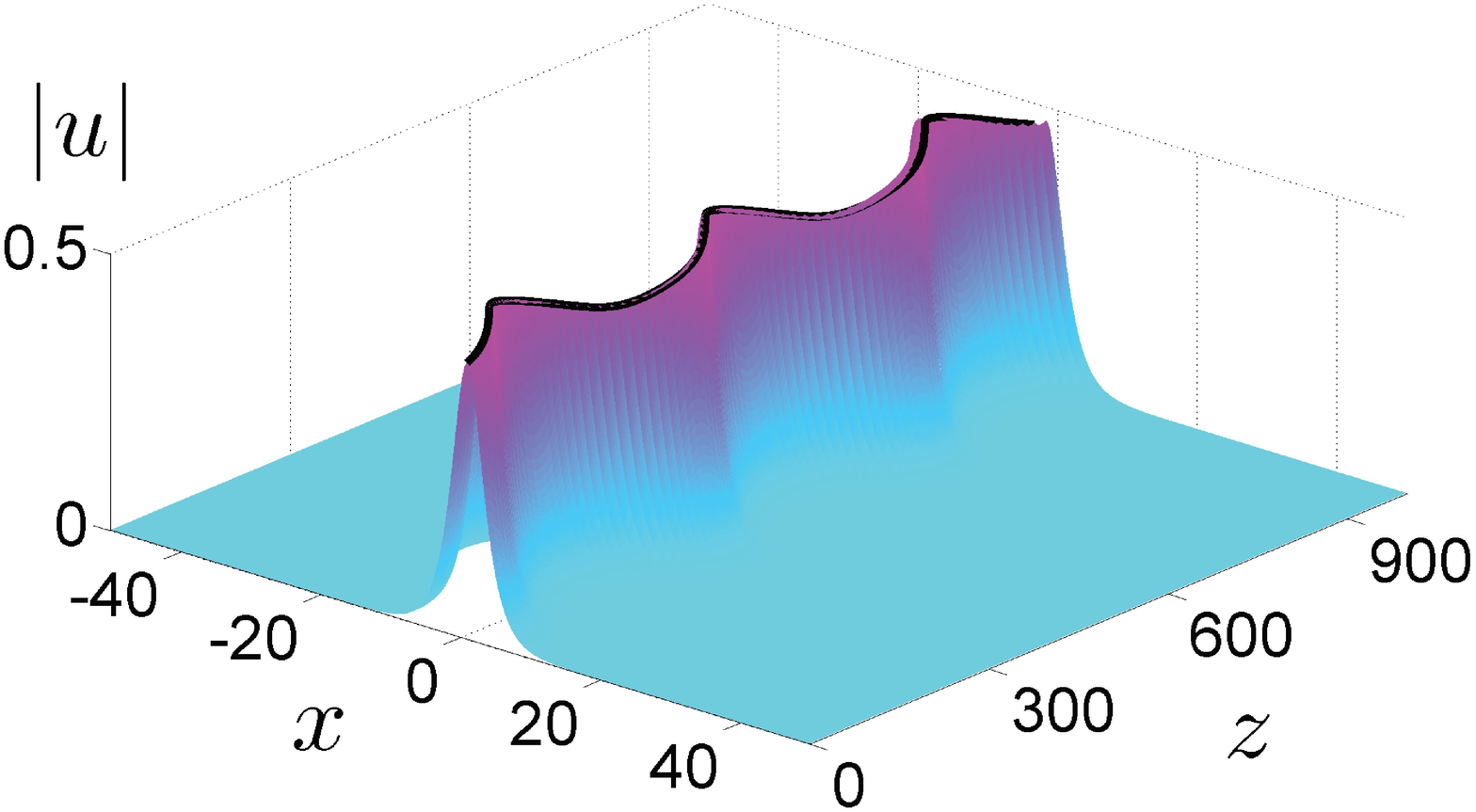}}}
	   \subfigure[]{\scalebox{\scl}{\includegraphics{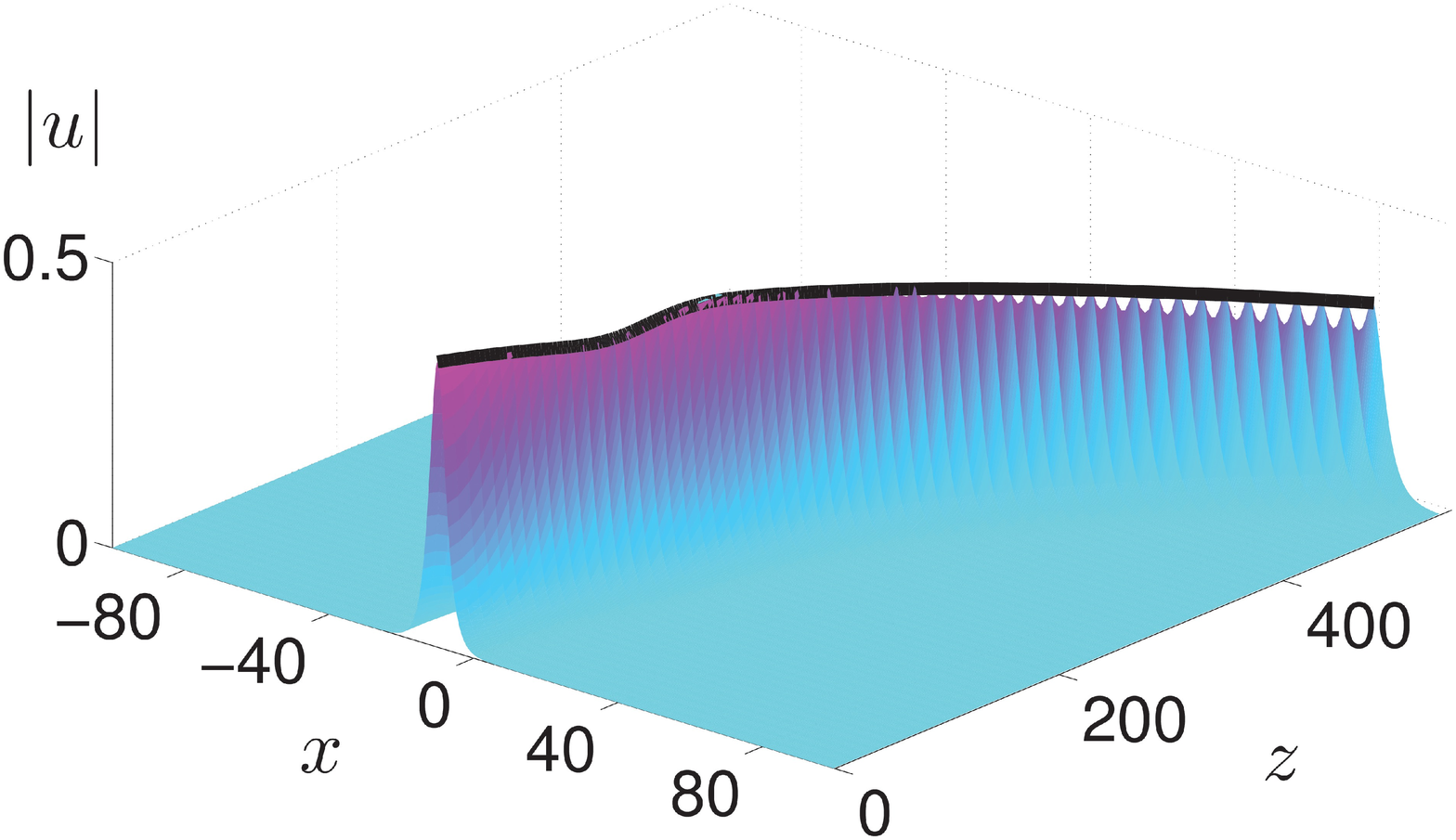}}}
\caption{Single hot-spot structure consisted of an amplifying part ($a_2=0.001$) of finite width $\Delta x=4$ in a lossy medium ($a_1=a_3=-0.0005$) for a refractive index profile fulfilling the condition (\ref{condition}) with $C=-1$ ($c_i=-a_i, i=1,2$). (a) Gain-loss $W(x)$ and refractive index $V(x)$ profiles [top]; Effective potential $U_{eff}(x_0)$ and mass variation rate $\Gamma(x_0)$ for a soliton of mass $m=1$ [bottom]. (b) Stationary propagation for initial soliton position $x_0=-2.65$ corresponding to the fixed point depicted by a thick dot in (a). (c) Trapped soliton oscillations for initial position $x_0=1$. (d) Travelling soliton propagation for initial position $x_0=-10$. The thick black lines depict results from the effective particle model.}
    \end{center}
\end{figure}

A typical realistic case with practical importance is a planar structure consisted of an amplifying part of finite width (hot-spot) in a lossy medium, with gain-loss and refractive index profiles fulfilling the condition (\ref{condition}) as shown in Fig. 2(a). For a soliton of mass $m=1$ and the parameters values of the specific structure, an asymmetric potential well can be formed as shown in Fig. 2(a). The fixed point located at $x_0=-6.25$ corresponds to stable stationary soliton propagation [Fig. 2(b)] whereas stable large amplitude oscillations (asymptotically evolving to the stationary soliton) can take place as a consequence of the condition for dynamic power balance, as shown in Fig. 2(c). Since the potential well has not an infinite depth, initial soliton conditions corresponding to untrapped dynamics result in travelling solitons of continuously decreasing mass, as illustrated in Fig. 2(d). It is worth noticing that the trapping conditions depend on both the parameters of the structure and the soliton mass, so that in each structure solitons having a mass below a critical value cannot be trapped. This fact results from the interplay of the two spatial scales, namely the soliton width ($\sim m^{-1}$) and the amplifying part width ($\Delta x$) as well as the relative magnitude of the gain and loss coefficients and is reflected in the effective particle model as a bifurcation of the fixed point corresponding to the local minimum of the effective potential. The spatial width and the depth of the potential well, for a given soliton mass value, determine the range of initial positions and angles of incidence (velocities) for soliton trapping and stability.

\begin{figure}[h]
   \begin{center}
  	   \subfigure[]{\scalebox{\scl}{\includegraphics{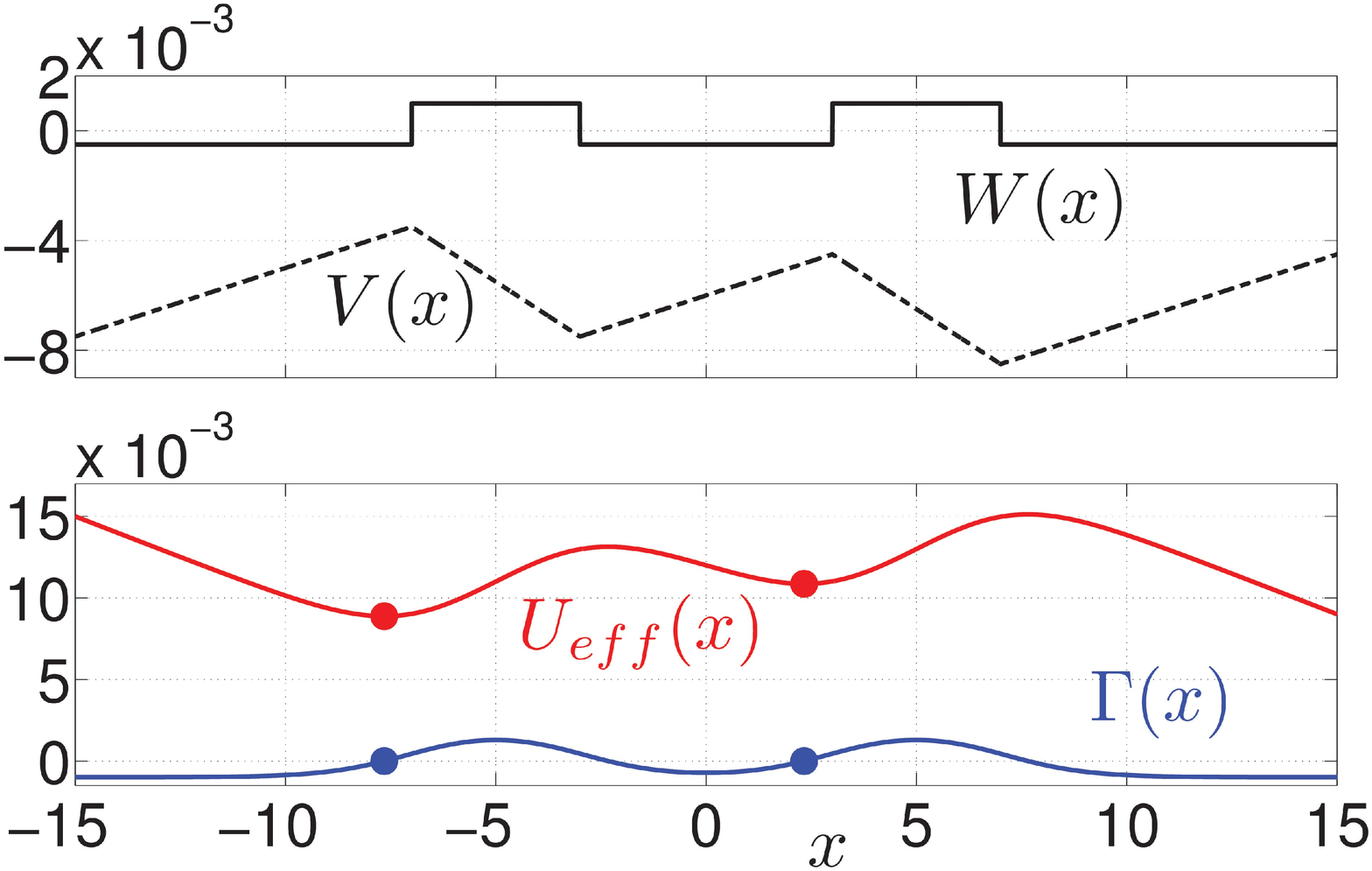}}}
	   \subfigure[]{\scalebox{\scl}{\includegraphics{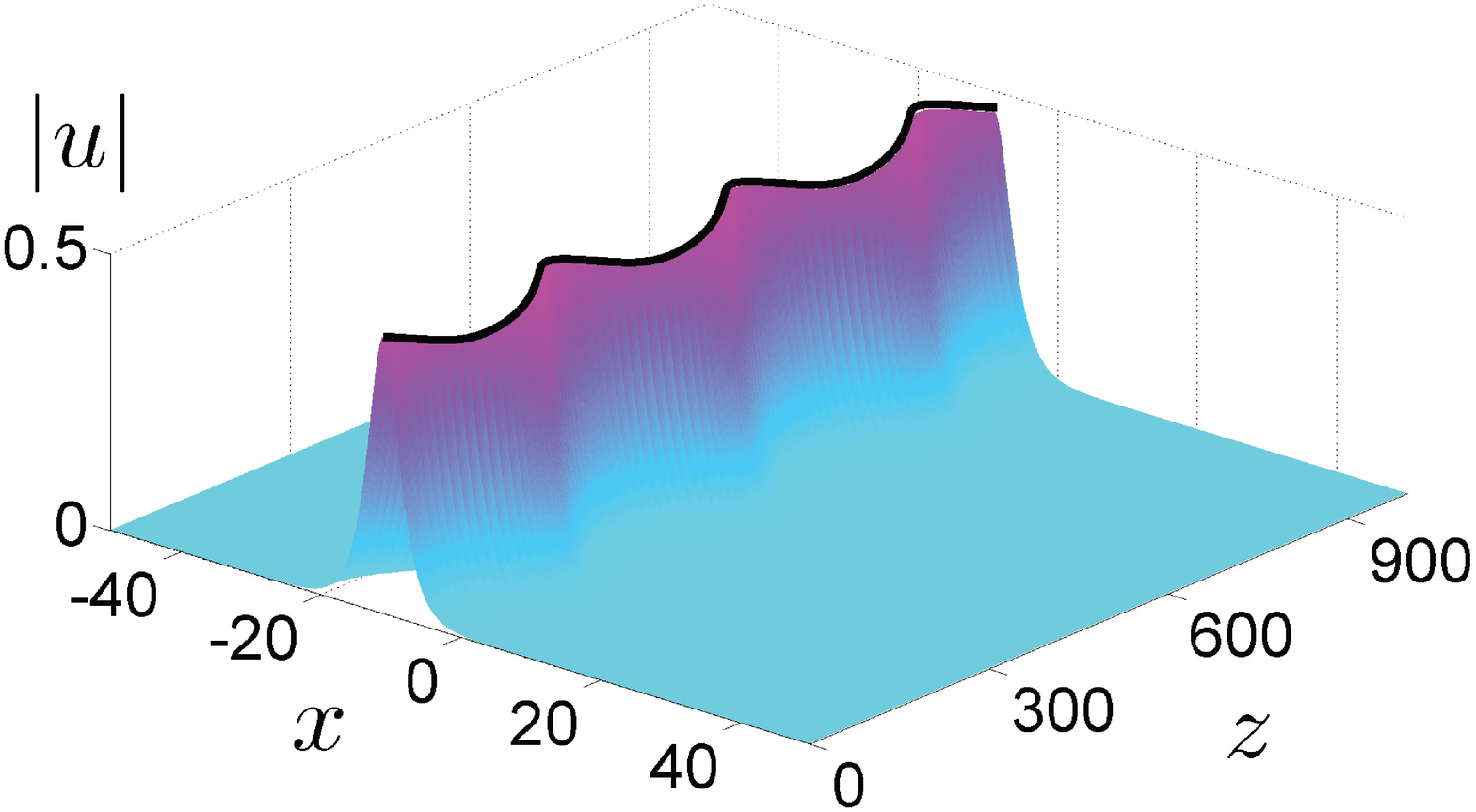}}}\\
	   \subfigure[]{\scalebox{\scl}{\includegraphics{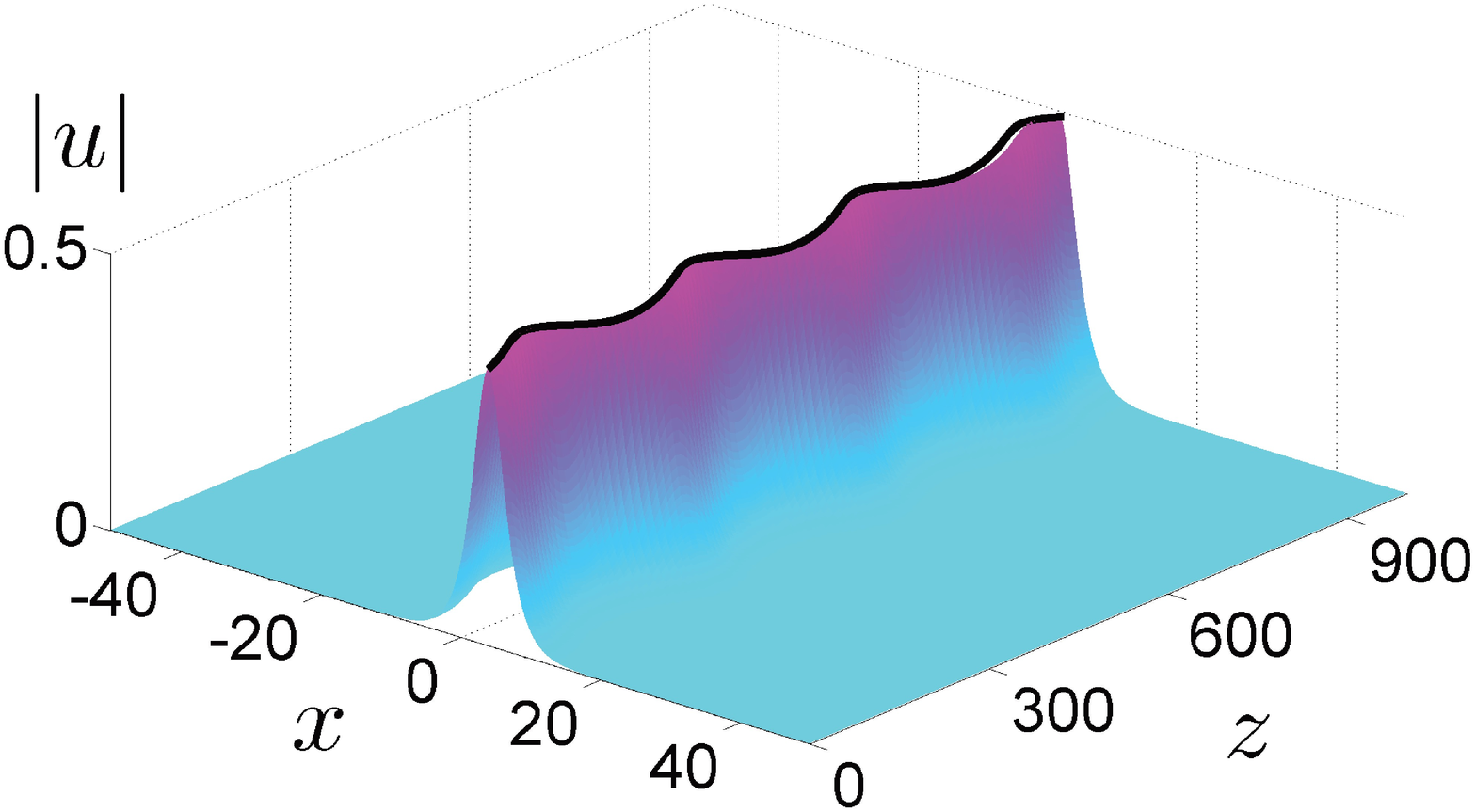}}}
	   \subfigure[]{\scalebox{\scl}{\includegraphics{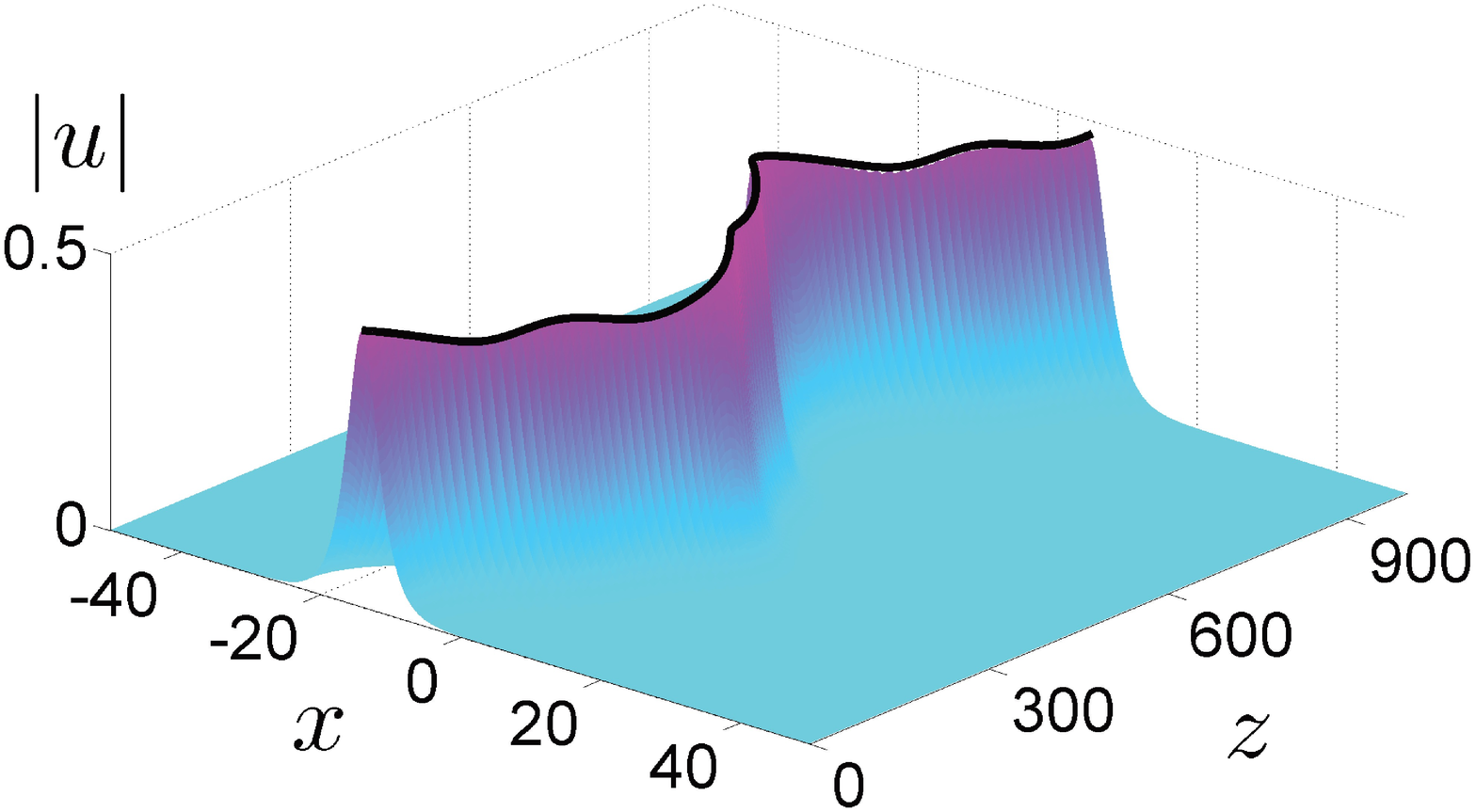}}}
\caption{Double hot-spot structure consisted of two amplifying parts ($a_2=a_4=0.001$) of finite width $\Delta x=4$ located at $x_c=\pm5$ in a lossy medium ($a_1=a_3=a_5-0.0005$) for a refractive index profile fulfilling the condition (\ref{condition}) with $C=-1$ ($c_i=-a_i, i=1,2$). (a) Gain-loss $W(x)$ and refractive index $V(x)$ profiles [top]; Effective potential $U_{eff}(x_0)$ and mass variation rate $\Gamma(x_0)$ for a soliton of mass $m=1$ [bottom]. (b) Trapped soliton oscillations in the left potential well for initial position $x_0=-11$. (c) Trapped soliton oscillations in the right potential well for initial position $x_0=4$. (d) Extended trapped soliton oscillations for initial position $x_0=-14$. The thick black lines depict results from the effective particle model.}
    \end{center}
\end{figure}

The case of a structure with two hot-spots is considered in Fig. 3. As expected, the increased complexity of the structure results in richer dynamics and trapping capabilities. In such case, we can have two fixed points, as shown Fig. 3(a). Therefore, under dynamic power balance conditions, trapping and stable soliton oscillations can occur either on the left potential well [Fig. 3(b)] or on the right one [Fig. 3(c)]. Moreover, for appropriate initial conditions, extended stable  oscillations can take place in the region above the two potential wells, for effective particle energy above the left and below the right local maximum of the effective potential [Fig. 3(d)]. The existence and bifurcations of the two fixed points again depend on the soliton mass and the parameters of the structure, so that we can have two, one or zero fixed points for a given soliton mass value.

\begin{figure}[h]
   \begin{center}
  	   \subfigure[]{\scalebox{\scl}{\includegraphics{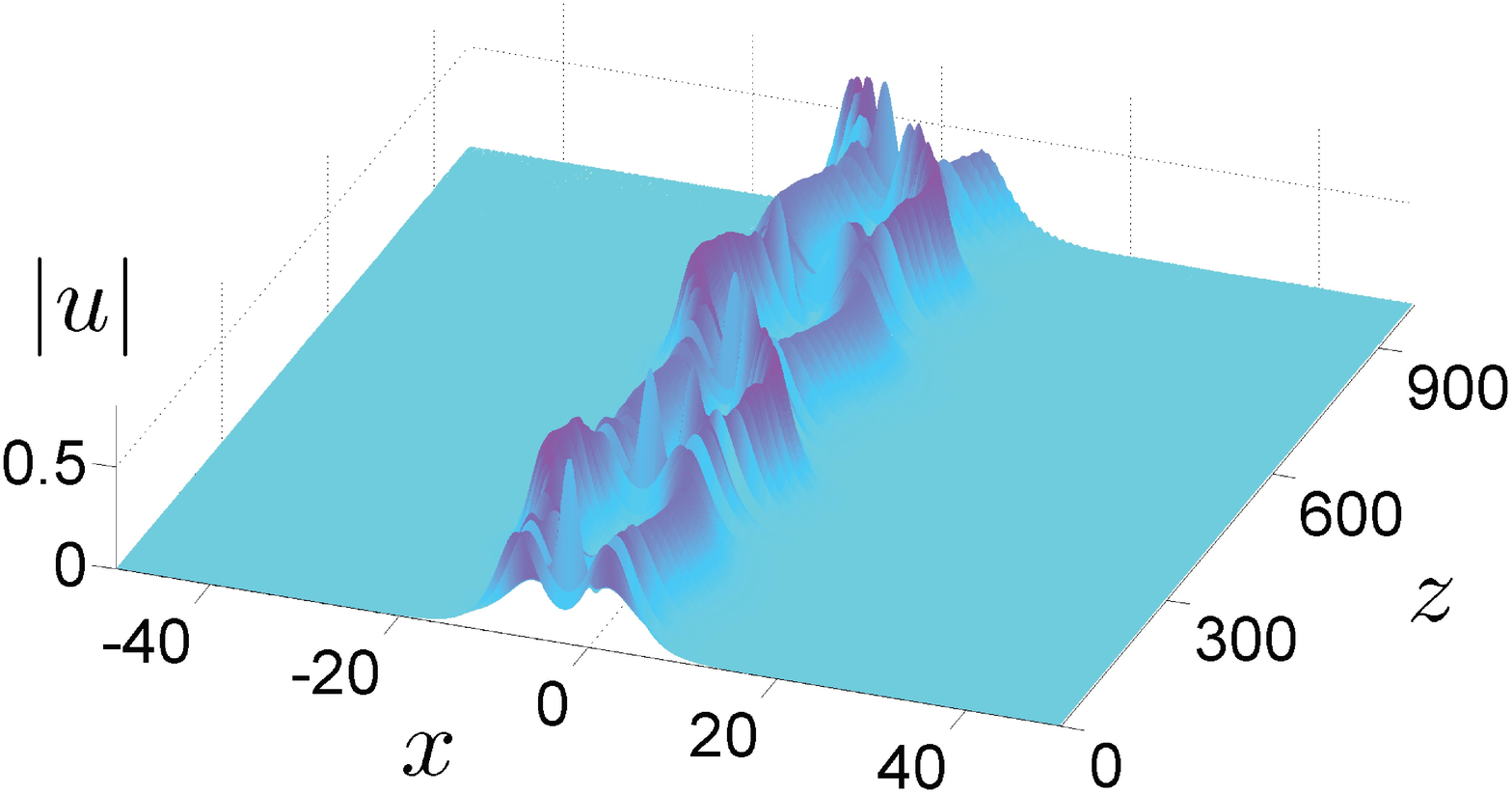}}}
	   \subfigure[]{\scalebox{\scl}{\includegraphics{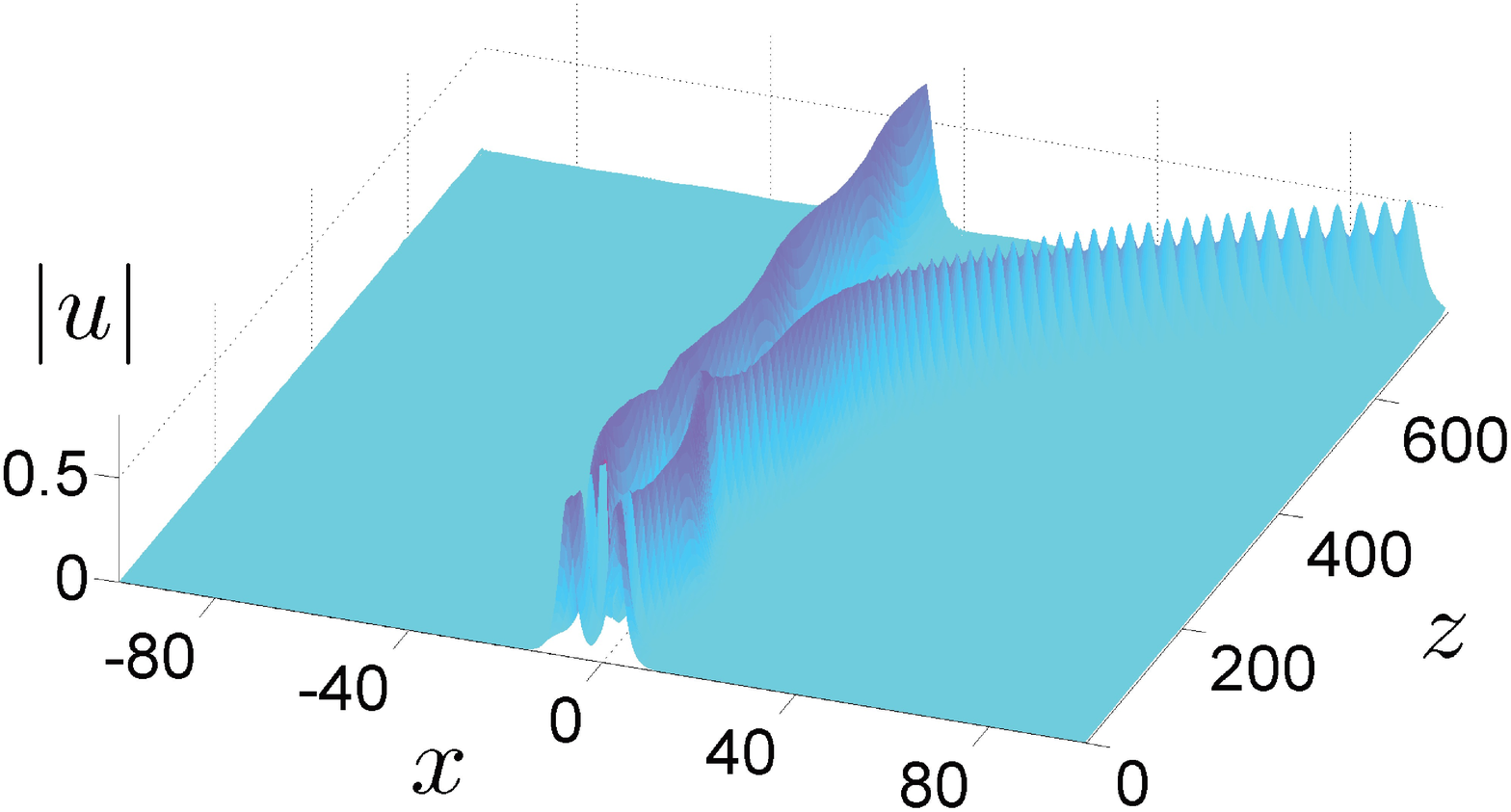}}}
\caption{(a) Oscillatory interaction of two solitons of mass $m=1$ and initial positions $x_0=-7.65$ and $x_0=2.33$ corresponding to the two fixed points depicted in Fig. 3(a). (b) Interaction of two solitons of mass $m=1.5$ and initial positions $x_0=-7.46$ and $x_0=2.54$ corresponding to the respective fixed points; the left soliton remains trapped whereas the right soliton is detrapped.}
    \end{center}
\end{figure}

Robust coexistence of two solitons trapped in the two different potential wells is shown in Fig. 4(a) for soliton mass $m=1$. Although initially located at the corresponding fixed points, solitons oscillate due to mutual interactions depending strongly both on the soliton width (mass) and the distance between the two hot-spots. Different interaction scenarios are possible as in the case of two solitons of higher mass ($m=1.5$) as shown in Fig. 4(b) where, although both solitons are launched at the corresponding fixed points, under interaction the right one is detrapped and travels with continuously decreasing mass whereas the other is stably trapped in its potential well. A numerous list for interaction scenarios, including solitons of different masses, can be considered which can be very interesting in terms of light control applications. It is worth emphasizing that it is the fulfillment of the dynamic power balance condition between the refractive index and the gain-loss profiles that allows for stable soliton dynamics and mutual interactions that could not take place either in the absence or in the inappropriateness of the trapping potential.

\section*{Conclusions} 
The fundamental problem of power balance of a nonlinear wave in a photonic structure with unbalanced gain and loss has been addressed. Sufficient conditions between the refractive index and gain-loss profiles have been derived for dynamic power balance of soliton propagation. In contrast to static power balance, that ensures only the existence of a fixed point corresponding to stationary soliton propagation, the dynamic power balance ensures the stability of the fixed point solution, allowing for stable propagation for a wide range of initial soliton positions and velocities, which is crucial for realistic applications. The analysis has been based on a simple effective particle model providing, not only the sufficient conditions, but also intuitive understanding of the complex soliton dynamics and being in remarkable agreement with the full model. The concepts and results of the dynamic power balance, illustrated here for simplicity only for piecewise constant gain-loss profiles, are so general that can be directly applied to any type of gain-loss profiles, nonlinear refractive index and nonlinear gain-loss inhomogeneities, as well as two-dimensional photonic structures.

\acknowledgments{This work has been partially supported by the Research Project NWDCCPS implemented within the framework of the Action ``Supporting Postdoctoral Researchers'' of the Operational Program ``Education and Lifelong Learning'' (Action's Beneficiary: General Secretariat for Research and Technology), and is co-financed by the European Social Fund (ESF) and the Greek State.}


\end{document}